\documentclass[preprint,showpacs]{revtex4}
\usepackage{graphicx}
\def\l{\langle}
\def\r{\rangle}

\begin{document}

\title{
Monte Carlo study of the antiferromagnetic three-state Potts model \\
with staggered polarization field on the square lattice}

\author{Yutaka Okabe}
\email{okabe@phys.metro-u.ac.jp}
\affiliation{
Department of Physics, Tokyo Metropolitan University,
Hachioji, Tokyo 192-0397, Japan
}

\author{Hiromi Otsuka}
\affiliation{
Department of Physics, Tokyo Metropolitan University,
Hachioji, Tokyo 192-0397, Japan
}

\date{\today}

\begin{abstract}
Using the Wang-Landau Monte Carlo method, we study 
the antiferromagnetic (AF) three-state Potts model 
with a staggered polarization field on the square lattice. 
We obtain two phase transitions; one belongs to the ferromagnetic 
three-state Potts universality class, and the other to the Ising 
universality class.  The phase diagram obtained is quantitatively 
consistent with the transfer matrix calculation. 
The Ising transition in the large nearest-neighbor interaction limit 
has been made clear by the detailed analysis of 
the energy density of states.  
\end{abstract}


\maketitle

\section{Introduction}

The $q$-state Potts model is one of the basic models for studying 
phase transitions \cite{Pott52}.  The properties of 
the antiferromagnetic (AF) Potts models are more complex than 
those of the ferromagnetic (F) ones.  The phase transitions of 
the AF Potts models depend heavily on the number of states $q$, 
the details of lattice structure, etc. 
The AF three-state ($q=3$) Potts model 
on a square lattice exhibits a second-order transition at $T$=0 
with the Gaussian criticality \cite{Lena67,Baxt70}.  
While this model with only the nearest-neighbor (NN) interactions 
has been studied in detail 
\cite{Lena67,Baxt70,Nijs82,Burt97,Wang89,Sala98,Ferr99,Card01}, 
the effect of the next-nearest-neighbor (NNN) interactions 
has many interesting problems to explore \cite{Nijs82,Otsu04}. 
Quite recently, the present authors have studied the square-lattice 
AF three-state Potts model with a staggered polarization field 
\cite{Otsu04}.  By the use of the exact diagonalization calculation 
of the transfer matrix and the phenomenological renormalization-group 
analysis, two types of phase transitions have been discussed 
in connection with a field theoretical argument. 
The crossover behavior from the AF to the F three-state Potts criticality, 
which was proposed by Delfino \cite{DelfNATO}, 
has been confirmed. 

The exact diagonalization calculation and the Monte Carlo simulation 
play a complementary role in the numerical study.  The energy levels 
obtained by the exact diagonalization are highly accurate, but 
the tractable size is limited.  On the other hand, although 
the statistical errors are unavoidable because of the sampling 
process, the Monte Carlo simulation can deal with larger systems.  
Moreover, the latter can easily study the spin configuration, 
the probability distribution of the order parameter, etc. 
Recently, several attempts have been proposed for 
the Monte Carlo algorithms to directly calculate the energy 
density of states (DOS), such as the multicanonical method 
\cite{berg91,Lee93}, and the Wang-Landau method \cite{wl}. 

In this paper, we study the square-lattice AF three-state Potts 
model with a staggered polarization field by using 
the Wang-Landau Monte Carlo method.  We calculate the energy DOS 
precisely, and study the phase transitions of the model. 
Using the finite-size scaling (FSS), we investigate the critical 
properties.  In Sec.~\ref{model}, the model and the calculation 
method are described.  
The results for phase transitions are given in Sec.~\ref{result}.  
The final section is devoted to a summary and discussions.

\section{Model and calculation method}
\label{model}

\begin{figure}
\includegraphics[width=0.9\linewidth]{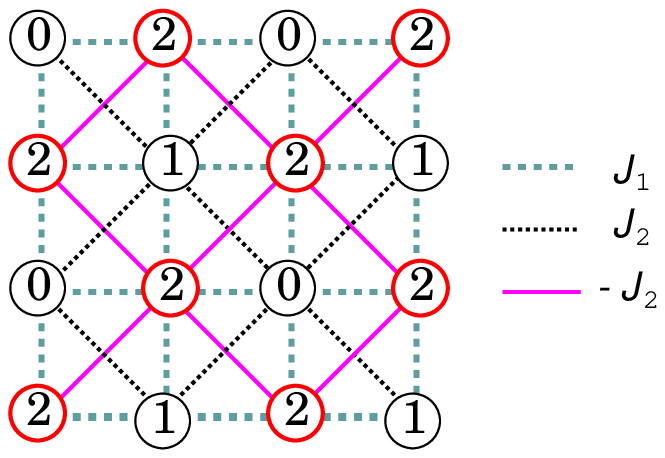}
\caption{Schematic illustration of the AF three-state Potts model 
with the staggered polarized field.  The sites belonging to the sublattice 
$\Lambda_+$ ($\Lambda_-$) are denoted by red thick (black thin) circles. 
The NNN interactions 
on the sublattice $\Lambda_+$ ($\Lambda_-$) are F (AF). 
An example of the ground-state configuration is also given. 
}
\label{interaction}
\end{figure}

We treat the AF three-state Potts model with the staggered 
polarization field on the square lattice $\Lambda$, 
whose Hamiltonian is given as
\begin{equation}
  H = J_1 \sum_{\l j,k\r} \delta_{\sigma_j,\sigma_{k}}
    - J_2 \sum_{[j,k]} (-1)^j \delta_{\sigma_j,\sigma_{k}}. 
   \label{eq_Hami1}
\end{equation}
Here, $\sigma_j = 0, 1, 2$, and $J_1, J_2 >0$. 
The first sum is performed over the whole NN pairs $\l j,k\r$, 
and the second sum over the whole NNN pairs $[j,k]$.  
Here, $(-1)^j = \pm 1$ for $j$ in the even (odd) sublattice $\Lambda_\pm$.
The second term is the staggered polarization field term, 
which breaks the sublattice symmetry.  
The NN and NNN interactions are schematically shown in Fig.~\ref{interaction} 
for convenience.  We also give an example of the ground-state configuration 
there. 

In order to obtain precise numerical information 
we use the Wang-Landau Monte Carlo algorithm \cite{wl} 
because of the following reasons.  This type of extended ensemble 
methods do not suffer from the problem of the critical slowing down 
near the second-order phase transitions.  The cluster algorithms 
are not so efficient for the systems with NNN couplings.  
We use the information of the energy DOS for the 
detailed study of the phase transitions. 
In the Wang-Landau method, a random walk in energy space 
is performed with a probability proportional 
to the reciprocal of the DOS, $1/g(E)$, which results in 
a flat histogram of energy distribution.  
Since the DOS is not known {\it a priori}, 
it is iteratively updated as 
\begin{equation}
\label{mod}
 \ln g(E) \rightarrow \ln g(E) + \ln f, 
\end{equation}
every time a random walker visits a state with energy $E$.  
A large modification factor $f$ is introduced 
to accelerate the diffusion of the random walk 
in the early stage of the simulation, 
and it is gradually reduced to unity by checking the `flatness' 
of the energy histogram.  We set the final value of ($\ln f$) 
as $10^{-8}$ following the original paper by Wang and Landau \cite{wl}, 
and then measure the energy dependence of a quantity $Q$, that is, $Q(E)$. 
In measuring $Q(E)$, the DOS $g(E)$ is fixed as the final one.
Finally we calculate the thermal average $\l Q \r_T$ at the temperature $T$ 
using the energy DOS and the Boltzmann weight 
as 
\begin{equation}
 \l Q \r_T = \frac{\sum_E Q(E) g(E) e^{-E/T}}{\sum_E g(E) e^{-E/T}},
\label{thermal}
\end{equation}
where the Boltzmann constant has been included 
in the definition of $T$.

We make simulations for several sets of $J_2/J_1$; we treat 
the system sizes $N = L \times L$ up to $L=64$.  
For the energy range of the random walk we do not cover 
the whole possible energy space to save computation time. 
The lowest energy is taken as the ground-state energy, 
$E = - J_2 N$, but we set the highest energy as that 
takes the maximum DOS, in other words, the energy 
at the infinite temperature.  This energy is $(2/3) J_1 N$. 
We make the measurement for $4 \times 10^6$ Monte Carlo 
steps per spin after the final energy DOS is obtained 
using the Wang-Landau process.  
We perform 64 independent runs for each system size 
in order to get better statistics and to estimate statistical errors. 
We estimate the statistical errors from 64 independent calculations. 
They could be underestimated if there are systematic errors 
due to the Wang-Landau method. 

\section{Results}
\label{result}

\subsection{Specific heat}

\begin{figure}
\includegraphics[width=0.9\linewidth]{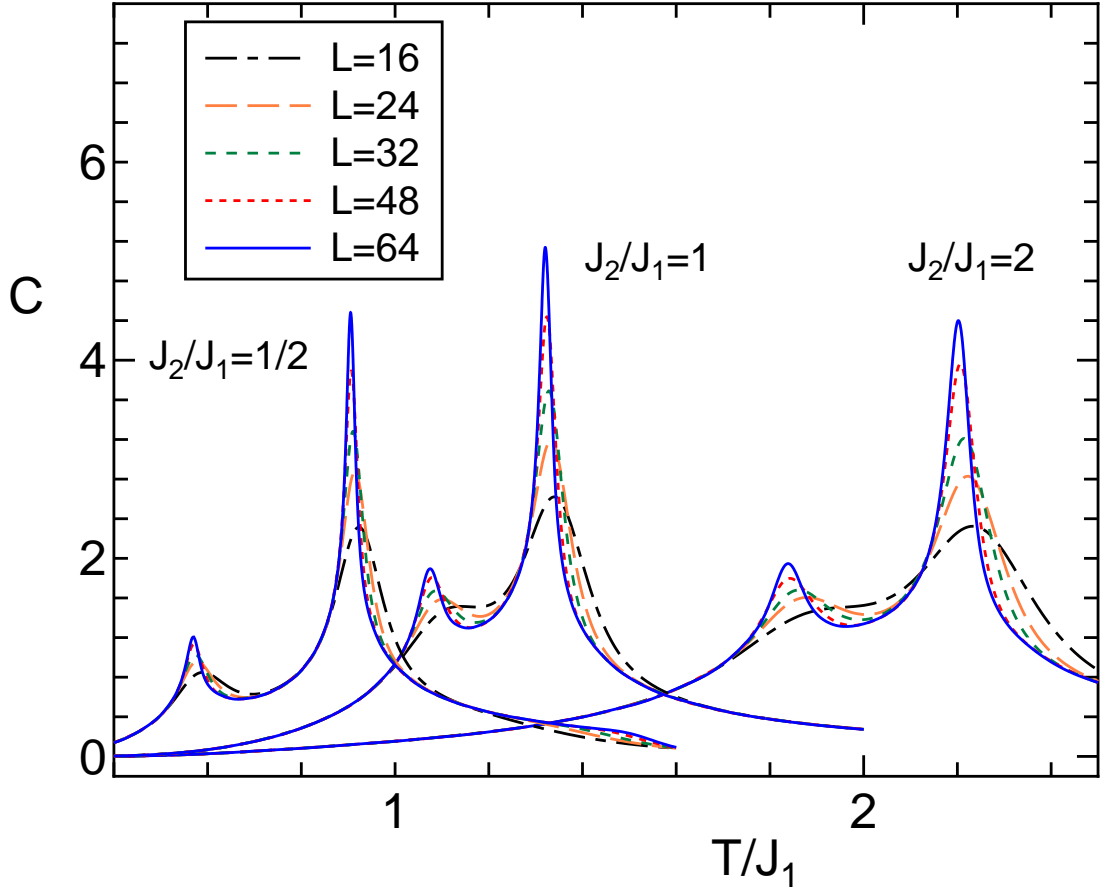}
\caption{The specific heat of the AF three-state Potts model 
with the staggered polarization field for $J_2/J_1$ = 1/2, 1, and 2.
The temperature is plotted in units of $T/J_1$.  The system sizes 
are $L$= 16, 24, 32, 48, and 64.  The statistical errors 
are within the width of lines. }
\label{Cv} 
\end{figure}

First, we present the data for the specific heat of the 
AF three-state Potts model with the staggered polarization field 
on the square lattice. 
We plot the temperature dependence of the specific heat 
for $J_2/J_1$ = 1/2, 1, and 2 in Fig.~\ref{Cv}. 
The temperature is denoted in units of $T/J_1$ 
from now on unless specified else.  
The system sizes are $L$=16, 24, 32, 48, and 64. 
The statistical errors are within the width of lines. 

We see two peaks in the specific heat for each $J_2/J_1$, 
which indicates the existence of two phase transitions clearly.  
The peak value for the high-temperature phase transition 
increases rapidly with the system size, which suggests 
a positive specific-heat exponent $\alpha$.  The peak value 
for the low-temperature transition, on the contrary, 
increases gradually with the size. 
It is difficult, however, to determine whether this increase is 
a logarithmic divergence or a power-law one with very small $\alpha/\nu$
for these system sizes.

\subsection{Order parameter}

\begin{figure}
\includegraphics[width=0.8\linewidth]{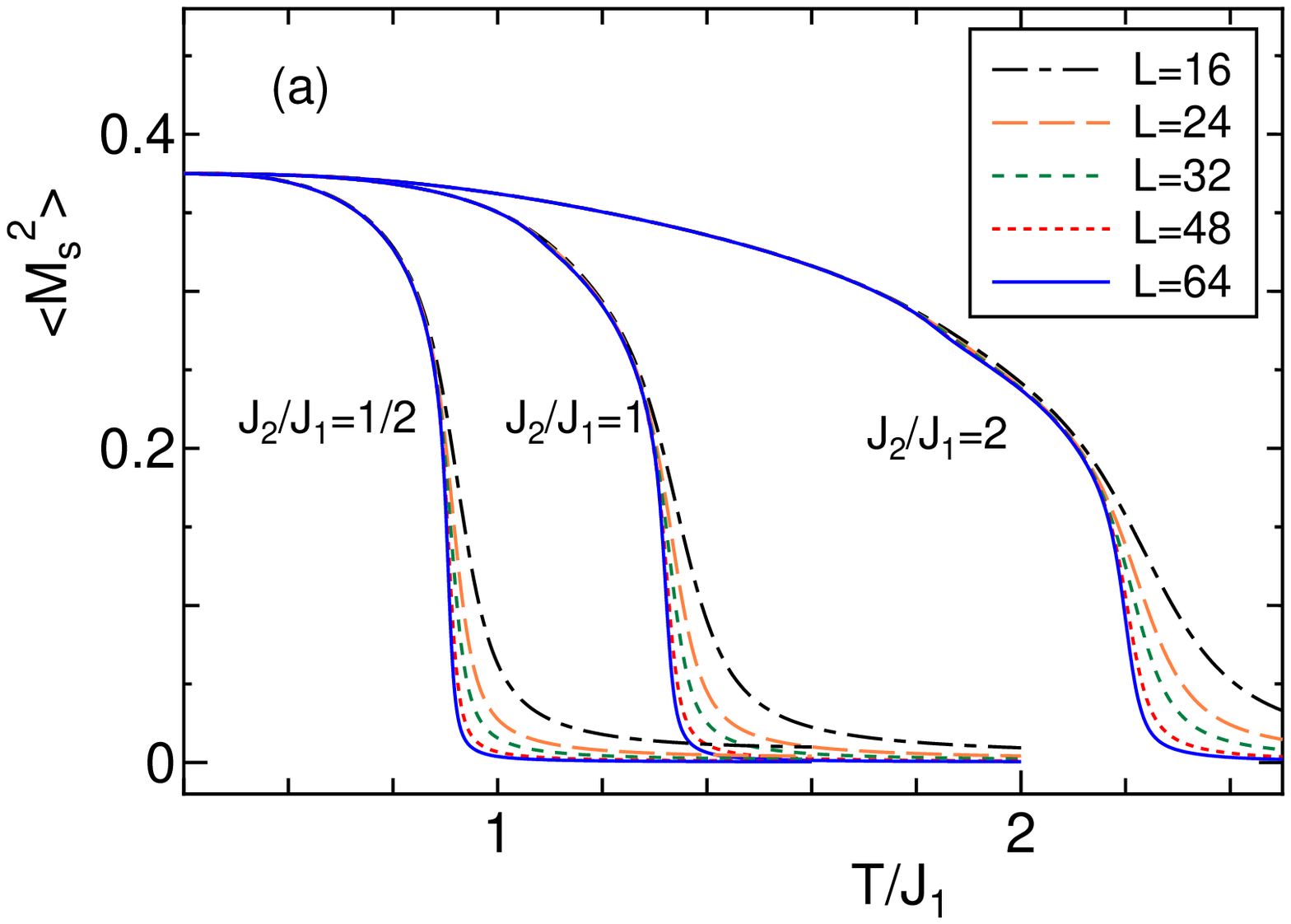}

\vspace{2mm}
\includegraphics[width=0.8\linewidth]{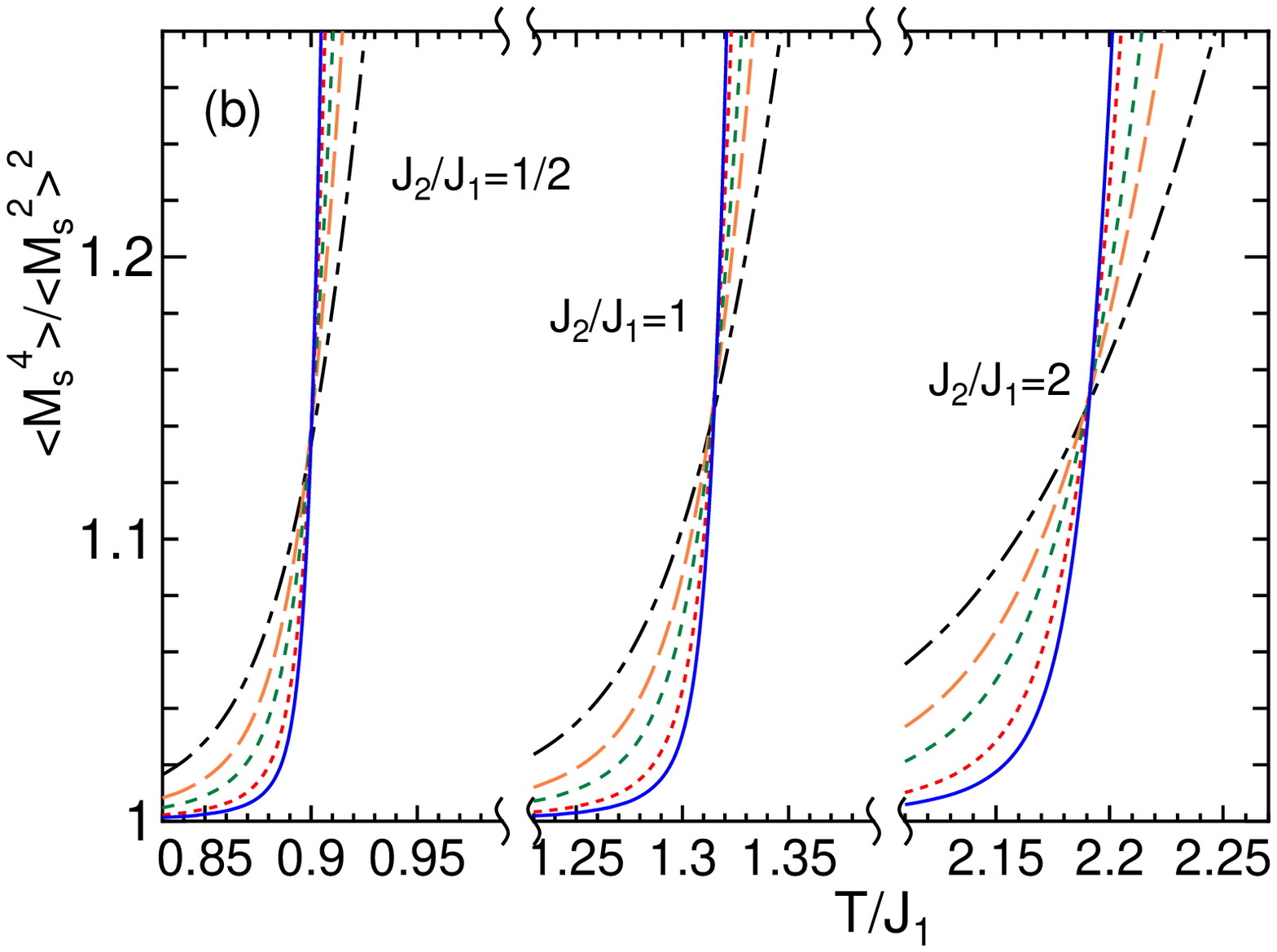}
\caption{The high-temperature order parameter (a) and 
the ratio of the its moments (b) of the AF three-state Potts model 
with the staggered polarization field for $J_2/J_1$ = 1/2, 1, and 2.
The temperature is plotted in units of $T/J_1$.  The system sizes 
are $L$= 16, 24, 32, 48, and 64.  The statistical errors 
are within the width of lines. }
\label{high} 
\end{figure}

To investigate the behavior of the phase transitions in more detail, 
we consider the order parameters.  For high-temperature phase 
transition, we look at the staggered magnetization for the 
AF Potts model, which is given by
\begin{equation}
 {M_{\rm s}}^2={S_0}^2+{S_1}^2+{S_2}^2,
\end{equation}
where
\begin{equation}
 S_\ell = \frac{1}{N} \sum_{j \in \Lambda} (-1)^j \, 
 \delta_{\sigma_j, \ell}, \quad \ell=0,1,2.
\end{equation}
The order parameter $M_{\rm s}$ takes non-zero values if 
the ${\bf S}_3$ symmetry associated with the global 
permutations of three Potts states is broken. 
At the ground state of the AF three-state Potts model 
with the staggered polarization field, which is shown 
in Fig.~\ref{interaction}, this order parameter 
takes the value of ${M_{\rm s}}^2 = 3/8$; 
we should note that the ground state is sixfold degenerate. 
The temperature dependence of $\l {M_{\rm s}}^2\r$
is plotted in Fig.~\ref{high}(a).  The value of $J_2/J_1$ is 
1/2, 1, and 2; for each $J_2/J_1$, the data for $L$=16, 24, 32, 48, 
and 64 are shown.  We see that $M_{\rm s}$ grows 
below the temperature that gives a peak of the specific heat, 
which indicates that $M_{\rm s}$ is an appropriate order 
parameter for the high-temperature phase transition. 
The F order develops on the sublattice $\Lambda_+$. 

\begin{figure}
\includegraphics[width=0.8\linewidth]{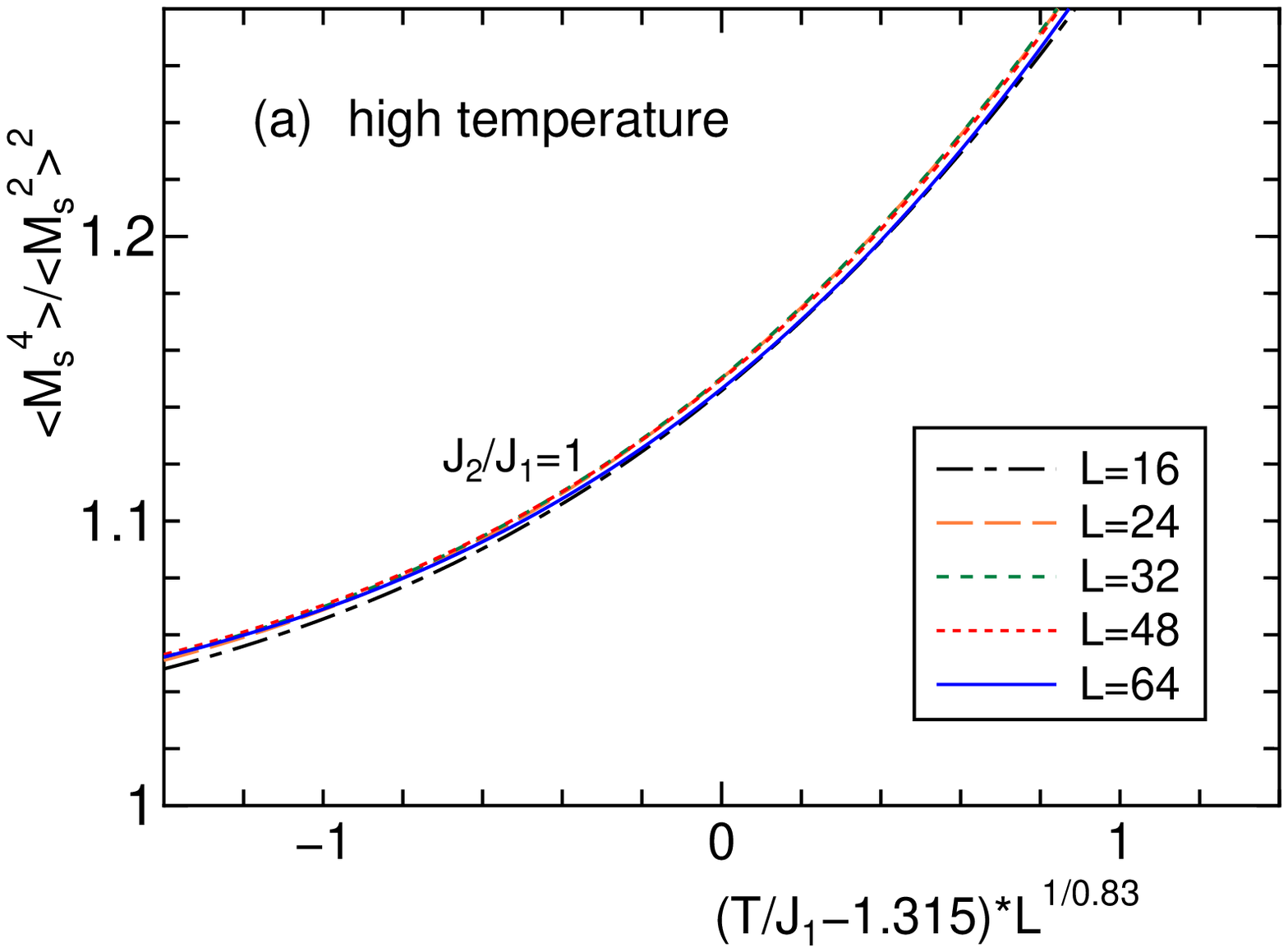}

\vspace{4mm}
\includegraphics[width=0.8\linewidth]{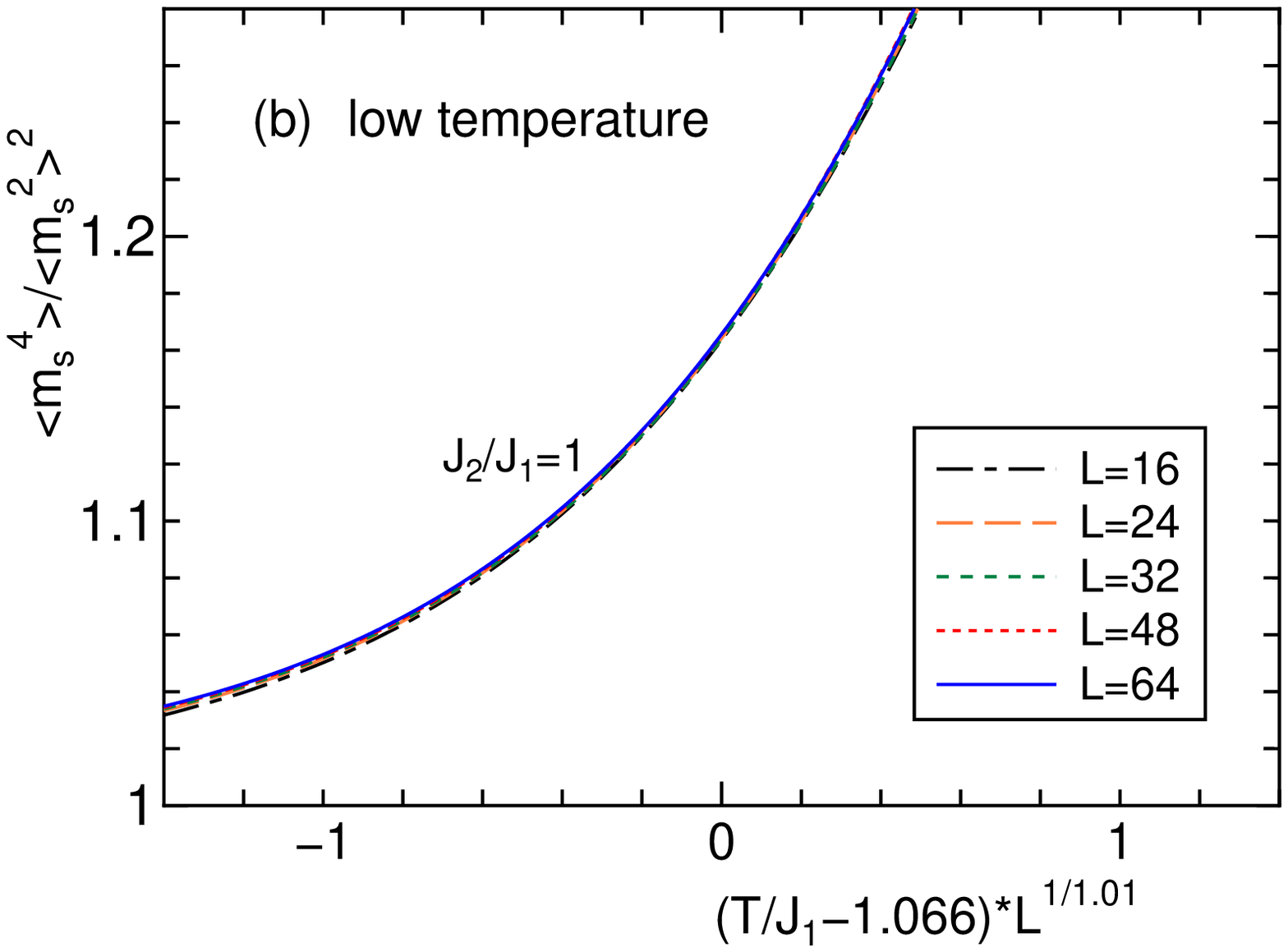}
\caption{Scaling plot of the moment ratios of the AF three-state Potts model 
with the staggered polarization field for $J_2/J_1$ = 1.
We plot both the high-temperature order parameter (a) and 
the low-temperature one (b). 
The temperature is plotted in units of $T/J_1$.  The system sizes 
are $L$= 16, 24, 32, 48, and 64.}
\label{FSS} 
\end{figure}

For the quantitative analysis of the phase transition, 
we use the moment ratio, $\l {M_{\rm s}}^4 \r/\l {M_{\rm s}}^2 \r^2$, 
which is essentially the same as the Binder ratio \cite{Binder}.
We plot the moment ratio for the high-temperature phase 
transition in Fig.~\ref{high}(b). 
Curves with different sizes cross at a single point if the 
corrections to FSS are negligible.  The crossings of 
our data are very good.  
From the crossing point, we estimate the critical temperature.  
We can also estimate the critical exponent $\nu$ by the FSS analysis,
\begin{equation}
 \l {M_{\rm s}}^4 \r/\l {M_{\rm s}}^2 \r^2 = f(tL^{1/\nu}),
\label{moment_ratio}
\end{equation}
where $t=(T-T_c)/T_c$.  
We show the FSS plot of the moment ratio for $J_2/J_1$=1, for example, 
in Fig.~\ref{FSS}(a).  We see a very good FSS.
We can also estimate the critical exponent $\beta/\nu$ 
using the FSS relation,
\begin{equation}
 \l {M_{\rm s}}^2 \r_{T=T_c} \sim L^{-2\beta/\nu}.
\end{equation}

We tabulate the estimates of $T_c$, $\nu$, and $\beta/\nu$ for 
the high-temperature phase transition in Table~\ref{table1}.  
The results for $J_2/J_1$ = 1/4 and 4 are also given there. 
The numbers in the parentheses denote the uncertainty 
in the last digits.  
We use the least-square fitting for the FSS estimates 
without considering the corrections to FSS.  
The statistical errors are estimated from 64 independent 
runs. 
We see from Table~\ref{table1} that $\nu$ is consistent with 
the F three-state Potts value of $\nu=5/6=0.833$.  
We also find that $\beta/\nu$ is compatible with 
the F three-state Potts value of $\beta/\nu=2/15=0.133$. 
The estimates of the moment ratio at $T=T_c$ are also given 
in Table~\ref{table1}.  This value is consistent with 
that for the F three-state Potts model, $1.16 \pm 0.01$ 
\cite{Tome}. 

Here, the corrections to FSS have not been considered 
explicitly for the estimate of the critical temperature 
and critical exponents.   The $J_2/J_1$ dependences 
of the estimated critical exponents and the moment ratio 
at $T=T_c$ are small, but the deviations from the values 
for the F three-state Potts model become larger for small $J_2/J_1$. 
This behavior is prominent for the moment ratio at $T=T_c$, 
which is more accurate than the critical exponents.  
This is consistent with the fact that 
the system on the sublattice $\Lambda_+$ is nothing but 
the F three-state Potts model for large $J_2/J_1$ limit, 
which determines the renormalization-group flow \cite{Otsu04}. 

Next consider the low-temperature phase transition. 
The ${\bf S}_3$ symmetry is already broken in
the sublattice $\Lambda_+$; we should consider the symmetry 
breaking within the sublattice $\Lambda_-$.  Then, we may consider 
the following quantity as the order parameter for 
the low-temperature phase transition: 
\begin{equation}
 {m_{\rm s}}^2={s_0}^2+{s_1}^2+{s_2}^2, 
\label{low_order}
\end{equation}
where
\begin{equation}
 s_\ell = \frac{1}{N} \sum_{k \in \Lambda_{-}} (-1)^k \, \delta_{\sigma_k, \ell}, 
 \quad \ell=0,1,2.
\end{equation}
We only look at the spins on the sublattice $\Lambda_-$.  
The sublattice $\Lambda_-$ forms a $\sqrt{2} \times \sqrt{2}$ square lattice, 
and we divide the sublattice $\Lambda_-$ into two further sublattices.  
For these further sublattices, we take $(-1)^k = \pm 1$ depending 
on the values of $k$, even or odd.  This order parameter essentially 
represents the AF Ising order in the sublattice $\Lambda_-$.  
At the ground state the low-temperature 
order parameter, Eq.~(\ref{low_order}), takes the value of 
${m_{\rm s}}^2 = 1/8$. 

\begin{figure}
\includegraphics[width=0.8\linewidth]{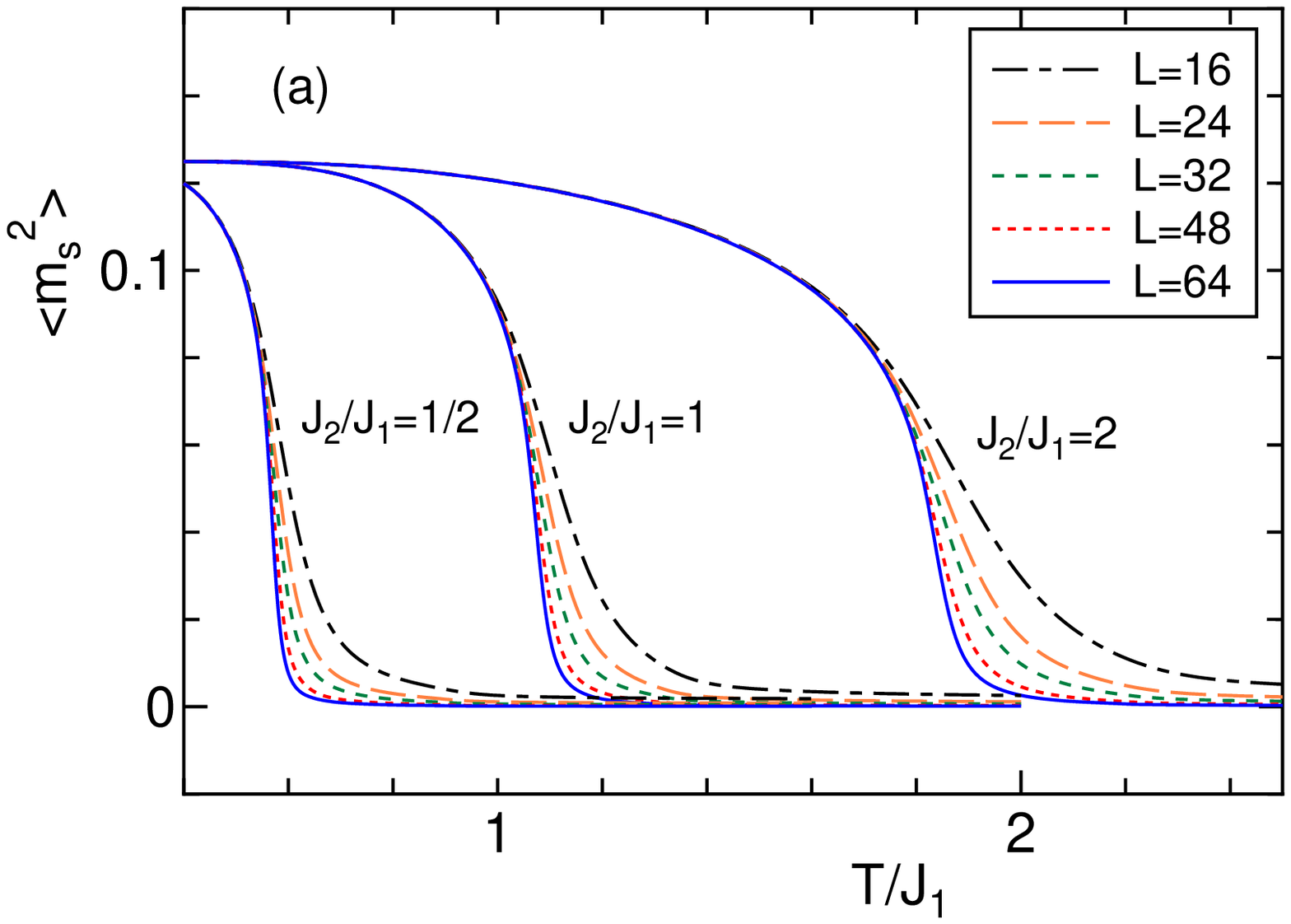}

\vspace{2mm}
\includegraphics[width=0.8\linewidth]{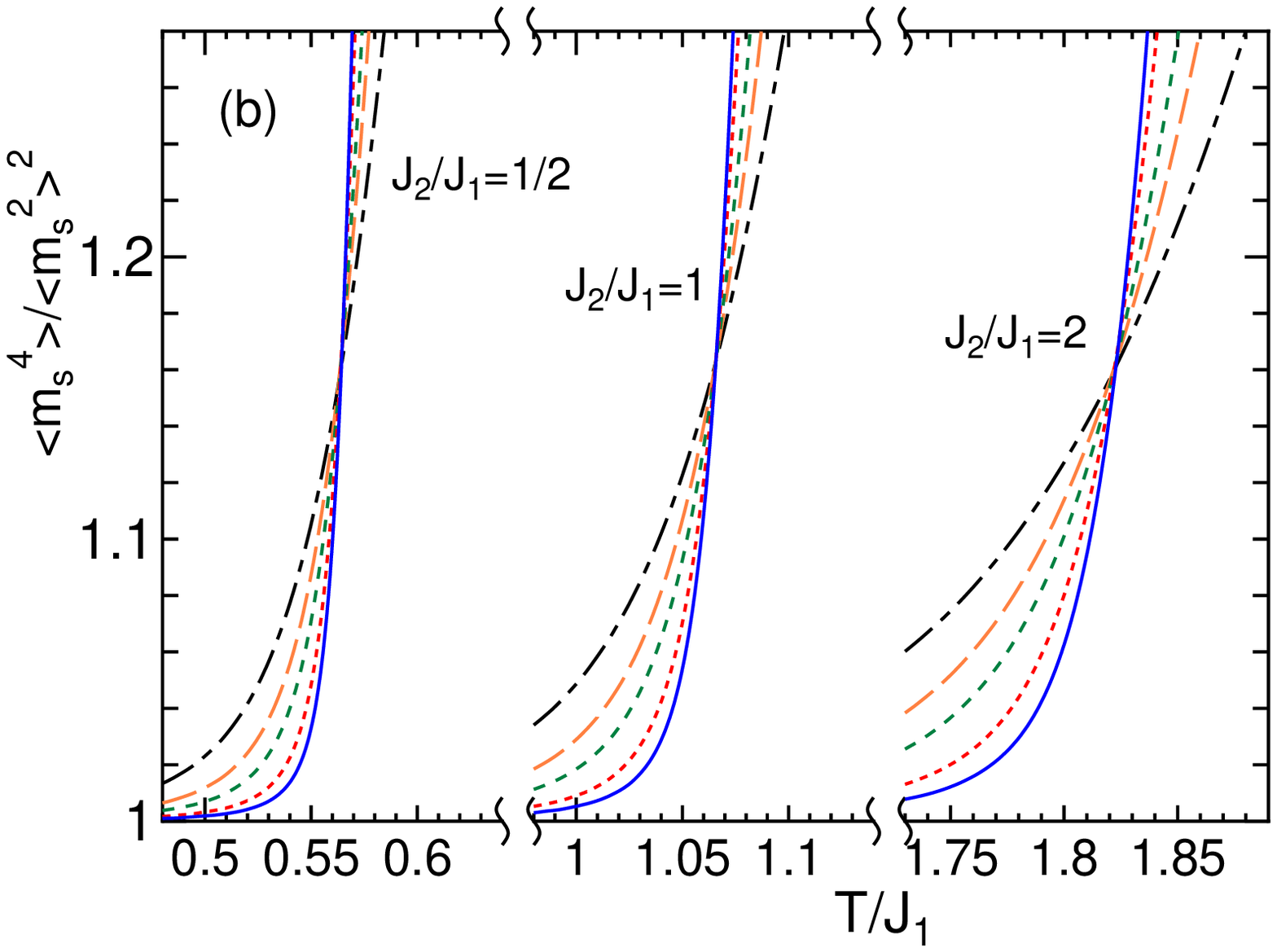}
\caption{The low-temperature order parameter (a) and 
the ratio of the its moments (b) of the AF three-state Potts model 
with the staggered polarization field for $J_2/J_1$ = 1/2, 1, and 2.
The temperature is plotted in units of $T/J_1$.  The system sizes 
are $L$= 16, 24, 32, 48, and 64.  The statistical errors 
are within the width of lines. }
\label{low}
\end{figure}

In Fig.~\ref{low}(a) we plot the order parameter of 
the low-temperature phase transition, Eq.~(\ref{low_order}).  
This time, $m_{\rm s}$ grows 
below the temperature that gives a lower peak of the specific heat; 
we see that $m_{\rm s}$ is an appropriate order parameter for the 
low-temperature phase transition.  We can consider 
the moment ratio associated with the low-temperature 
order parameter; $M_{\rm s}$ will be replaced by $m_{\rm s}$ 
in Eq.~(\ref{moment_ratio}). This moment ratio is 
plotted in Fig.~\ref{low}(b).  Again, we see the crossing 
of the curves with different sizes.  Using the FSS analysis, 
we estimate $T_c$, $\nu$, and $\beta/\nu$, and they 
are tabulated in Table~\ref{table1}.  The FSS plot of the moment ratio 
for $J_2/J_1$=1 is given in Fig.~\ref{FSS}(b), for example. The estimated 
exponents shown in Table~\ref{table1} suggest that the exponents $\nu$ 
and $\beta/\nu$ are consistent with the two-dimensional (2D) Ising values, 
1 and 1/8=0.125, respectively.  The moment ratio at $T=T_c$ is also 
given in Table~\ref{table1}.  The Ising value for the finite system 
with aspect ratio 1:1 is rigorously calculated as 1.1679229(47) 
\cite{Francesco}.  The coincidence with this value is very good. 

The $J_2/J_1$ dependences of the estimated critical exponents 
and the moment ratio at $T=T_c$ are very small, but the deviations 
from the Ising values become slightly larger for large $J_2/J_1$, 
which is the opposite direction from the case of 
the high-temperature transition.  

\subsection{Phase diagram}

\begin{figure}
\includegraphics[width=0.9\linewidth]{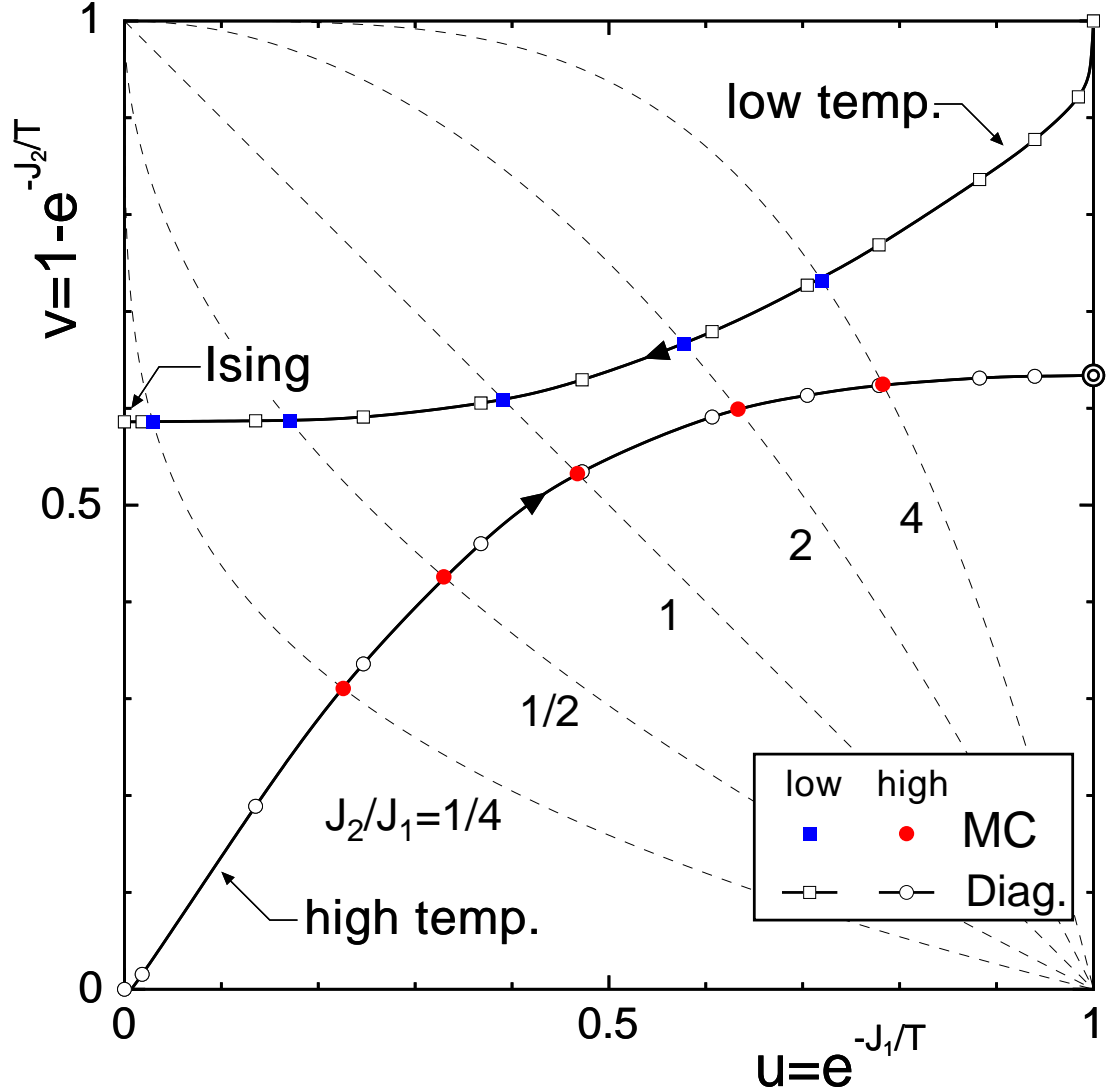}
\caption{Phase diagram of the AF three-state Potts model 
with the staggered polarization field. 
The filled circle (red) and filled square (blue) represent 
the estimate of high-temperature and low-temperature $T_c$'s 
in the present Monte Carlo study, 
whereas the open marks represent those by the 
transfer-matrix calculation \cite{Otsu04}.  
The F three-state Potts point $(u,v)=(1,(3-\sqrt{3})/2=0.6340)$ 
is shown by the double circle, and a point with 2D-Ising criticality 
on the $v$-axis 
is shown by the arrow.  
}
\label{phase} 
\end{figure}

From the list of the estimates of the two critical temperatures 
given in Table~\ref{table1} for $J_2/J_1$ = 1/4, 1/2, 1, 2, and 4, 
we discuss the phase diagram.  In Fig.~\ref{phase} 
we plot the phase diagram in the parameter space 
of $(u,v)=(e^{-J_1/T}, 1-e^{-J_2/T})$, which is the same 
as the previous transfer-matrix study \cite{Otsu04}. 
The trajectories of $J_2/J_1$=1/4, 1/2, 1, 2, and 4 are 
shown by dotted curves starting from $T=0$ [$(u,v)=(0,1)$] 
to $T=\infty$ [$(u,v)=(1,0)$].
The filled circle (red) and filled square (blue) represent 
the estimate of high-temperature and low-temperature $T_c$'s 
in the present study, respectively, and they are 
compared with the previous estimates by the 
transfer-matrix calculation \cite{Otsu04}, which are 
given by open marks.  
These two results are consistent with each other, 
which shows the reliability of both calculations.  
In the transfer-matrix calculation~\cite{Otsu04}, 
the characterization of the excitation levels was 
important.  In the present MC simulation, 
we have employed the method to calculate the DOS accurately, 
and have made the appropriate choice of the order parameters. 
With these careful treatments, we have obtained 
the accurate enough phase diagram.  

From the behavior of the corrections, we can see the renormalization-group 
flow of the phase boundary, which is consistent 
with Zamolodchikov's $c$-theorem \cite{Zamo86}.  
For the high-temperature phase transition, 
the flow starts from the Gaussian point $(u,v)=(0,0)$ to the 
F three-state Potts point $(u,v)=(1,(3-\sqrt{3})/2=0.6340)$, 
which is shown by the double circle.  
On the other hand,
for the low-temperature phase transition, the flow starts 
another Gaussian point $(u,v)=(1,1)$ to a point with 2D-Ising criticality 
on the $v$-axis, which is shown by the arrow in Fig.~\ref{phase}. 

This Ising transition in the large $J_1$ limit (on the $v$-axis) 
is a subtle problem.  In Ref.~\cite{Otsu04}, the inverse critical temperature 
was estimated as $J_2/T_c$ = 0.8820, 
which is slightly larger than the Ising value 
of $\ln(\sqrt{2}+1)$=0.8814 [$v=1-e^{-J_2/T}$=$2-\sqrt{2}$=0.5858].  

\subsection{Large NN interaction limit}

Here we examine the Ising transition in the large $J_1$ limit. 
If we consider only the NN interaction term in Eq.~(\ref{eq_Hami1}), 
the energies of the ground states and the first excited states 
are 0 and $J_1$, respectively.  The NNN interaction term takes 
the energy between $-J_2 N$ to $J_2 N$.   
Thus, for the case of $J_2/J_1 \le 1/(2N)$, 
we have to consider only the ground-state configuration 
for NN interactions.  
We make a Wang-Landau type MC simulation for our model 
with this restriction.  
That is, only the ground-state configurations for the 
NN AF three-state Potts model are allowed, and we calculate 
the energy DOS for the NNN interactions.

\begin{figure}
\includegraphics[width=0.9\linewidth]{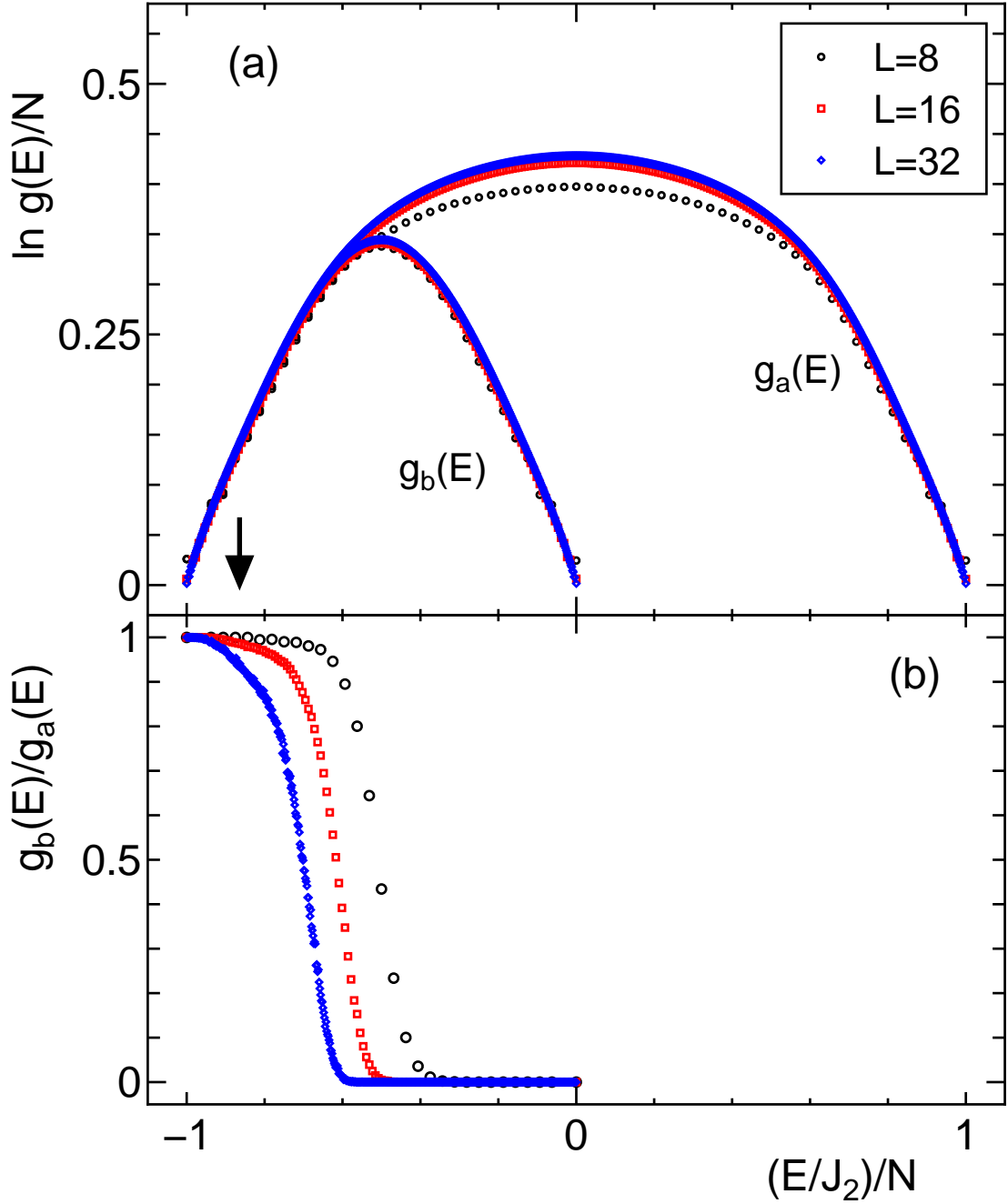}
\caption{(a) The energy DOS of the AF three-state Potts model with 
the staggered polarization field in the subspace of 
the ground states of the NN AF three-state Potts model, $g_a(E)$, 
and that of the pure Ising model, $g_b(E)$. 
The critical energy of the Ising transition point, 
$(E/J_2)/N$ = $- (2+\sqrt{2})/4=-0.8536$, is shown by an arrow. 
(b) The ratio of two DOS's, $g_b(E)/g_a(E)$.
}
\label{DOS} 
\end{figure}

\begin{figure}
\includegraphics[width=0.9\linewidth]{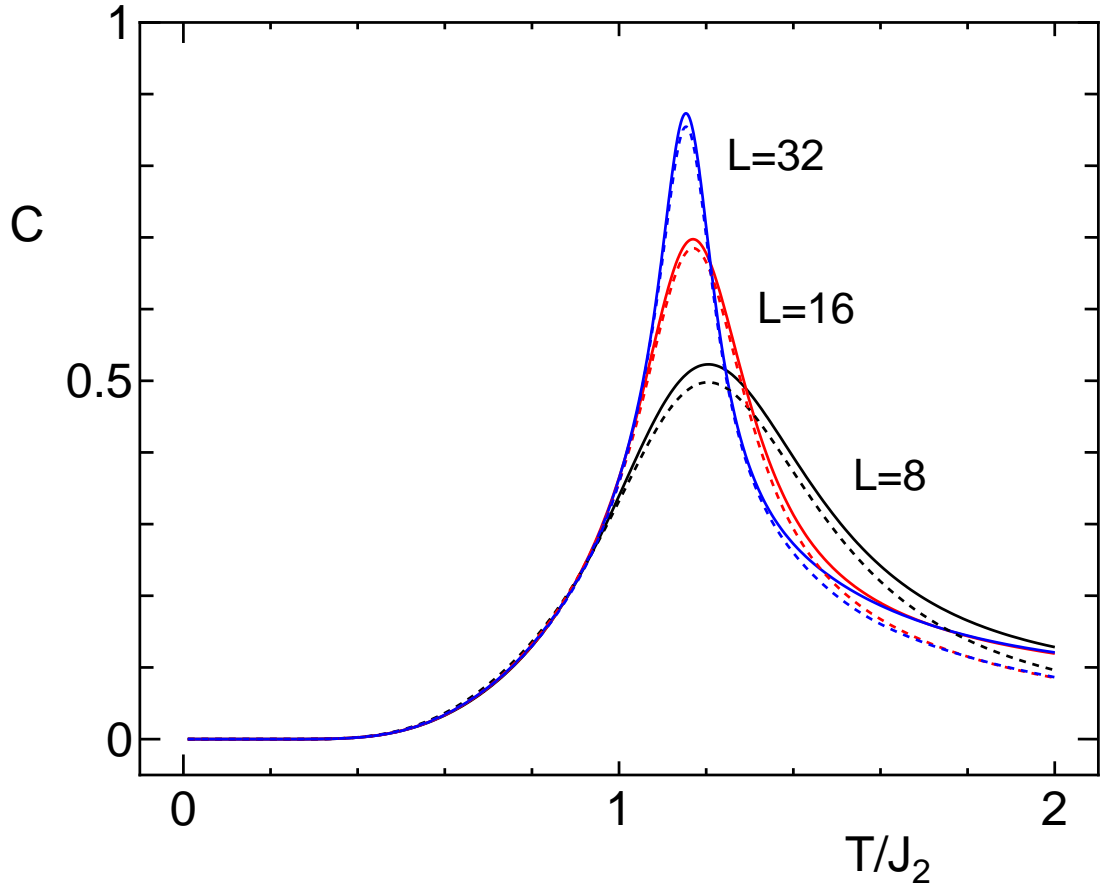}
\caption{Comparison of the specific heat between that for 
the present model in the subspace of the ground states of 
the NN AF three-state Potts model (solid line) and 
that for the pure Ising model (dotted line).}
\label{Cv2} 
\end{figure}

In Fig.~\ref{DOS}(a), we plot the DOS for this restricted model 
in the subspace of the ground states of the NN AF three-state 
Potts model, $g_a(E)$.  
If all the spins on the sublattice $\Lambda_+$ take one of 
the three states (the complete F three-state Potts order) 
and the spins on the sublattice $\Lambda_-$ take either one of 
the other two states, these spin configurations are parts of 
the ground states of the NN AF three-state Potts model, 
and are called the broken-sublattice-symmetry states 
in the study of the AF Potts models \cite{Banavar}.  
Then, the spins on the $\sqrt{2} \times \sqrt{2}$ sublattice $\Lambda_-$ 
are regarded as the AF Ising model with the NN interactions. 
The energy DOS of this pure Ising model, $g_b(E)$, is also shown 
in Fig.~\ref{DOS}(a). 
The extra three-times degeneracy due to three complete F states 
in the sublattice $\Lambda_+$ was taken into account. 
The phase transition is determined by the structure of the energy DOS 
near the Ising critical energy, $(E/J_2)/N$ = $- (2+\sqrt{2})/4=-0.8536$, 
which is shown by an arrow in Fig.~\ref{DOS}. 
In this region two DOS's look close when the logarithmic scale 
is used.  However, we understand the very small difference in 
the critical temperature from the structure of DOS. 
We plot the ratio of $g_b(E)/g_a(E)$ in Fig.~\ref{DOS}(b). 
Then, we find that this ratio becomes smaller for 
larger system size $L$.  This means that we cannot ignore 
contributions from the spin configurations which are 
not the pure Ising one.  Therefore, we may conclude that 
although this transition belongs 
to the Ising universality class, the critical temperature, 
which is not a universal quantity, is slightly modified 
from that of the pure Ising model.  Since the number of configurations 
are larger than that for the pure Ising model, the critical 
temperature $T_c/J_2$ becomes slightly lower than 
that for the pure Ising model.  
In Fig.~\ref{Cv2}, we compare 
the specific heat calculated from two DOS's, 
which give almost the same behavior but there is still a small 
difference. 

In the MC method to calculate DOS we obtain the relative ratio of DOS's 
at different energies $E_1$ and $E_2$, $g(E_1)/g(E_2)$.  
The absolute value of $g(E)$ can be obtained with other conditions.  
For the Ising model, as an example, the equation to give the total 
number of states, $\sum_E g(E)=2^N$, is such a condition.  
There is also a boundary condition if the ground-state degeneracy 
is known.  The ground state of the present model is 
sixfold degenerate; thus $g(E=-J_2N)$=6. 
Then, with this condition we can calculate the total number of states, 
$\sum_E g(E)$.  This value is nothing but the ground-state degeneracy 
for the NN AF three-state Potts model, because we restrict ourselves 
to the subspace of the ground states of the AF three-state Potts model.  
To confirm the reliability of our calculation, 
we calculate the normalized ground-state entropy of 
the AF three-state Potts model, $S/N = [\ln \sum_E g(E)] /N$, 
as a function of $L$ with this procedure, $S(1)/N$, 
and they are tabulated in Table~\ref{table2}.  
More direct way of obtaining the ground-state entropy 
is to calculate the ground-state DOS of the three-state 
Potts model with the condition of $\sum_E g(E)=3^N$.  This 
procedure was employed in the study of 
the three-dimensional AF Potts models \cite{Yama01}.  
The result of this direct way, $S(2)/N$, is also tabulated 
in Table~\ref{table2}.  
These two results coincide with each other completely. 
This indicates the effectiveness of the treatment 
of this subsection, that is, 
we have applied the Wang-Landau method to the present model 
with the restricted configurations.  
The normalized ground-state entropy for the AF three-state Potts model 
on the square lattice is exactly known as $\ln (4/3)^3/2 = 0.431523$ 
for the infinite size limit \cite{ent}.  
The extrapolation of the data of $S/N$ shown in 
Table~\ref{table2} as a polynomial of $1/N$ yields $0.43154+1.10 \times (1/N)$, 
which is consistent 
with the exact value within the statistical errors. 
We should mention that the ground-state entropy of the AF Potts model was 
extensively studied numerically by Shrock and Tsai \cite{Shrock}.

\section{Summary and discussions}

We have studied the square-lattice AF three-state Potts model 
with a staggered polarization field using the Wang-Landau MC method.
We have obtained two phase transitions, which belong to the ferromagnetic 
three-state Potts and Ising universality classes.  
We have confirmed the quantitative consistency of the phase diagram 
with the transfer-matrix study~\cite{Otsu04}. 
This consistency shows the reliability of both calculations, 
the previous transfer-matrix calculation and the present MC study. 
A special attention has been paid to the Ising transition 
in the large $J_1$ limit.  The origin of the slight difference 
of the transition point from the value of the pure Ising model 
has been made clear by the detailed analysis of the energy DOS.
As a check of the method, we have calculated the ground-state residual entropy 
for the AF three-state Potts model by two ways. 

The specific heat has singularities at the critical points as shown 
in Fig.~\ref{Cv}.  
The specific-heat amplitudes have information on the crossover 
behavior \cite{Cardy}.  We can study the Gaussian to F three-state Potts 
and the Gaussian to Ising crossovers by the $J_2/J_1$ dependence of 
the specific-heat amplitudes.  The detailed study on this crossover 
will be left to a separate study.  

We here make comments on the Wang-Landau method.  
In calculating the thermal average of $M_{\rm s}$ and $m_{\rm s}$, 
we have used the microcanonical average $M_{\rm s}(E)$ and $m_{\rm s}(E)$ 
as in Eq.~(\ref{thermal}).  If the energy DOS g(E) is not converged 
enough, it may cause a systematic error in calculating 
the thermal average.  The use of the joint DOS $g(E,Q)$ 
may help the convergence.  We have checked 
for smaller system sizes that the present calculations 
with the microcanonical average and those using the joint DOS 
give the same results within the statistical errors.
The second comment is as follows. 
We may employ a random walk in the space of two parameters, 
$\sum_{\l j,k\r} \delta_{\sigma_j,\sigma_{k}}$ 
and $\sum_{[j,k]} (-1)^j \delta_{\sigma_j,\sigma_{k}}$,  
instead of the single parameter of the total energy $E$.  
Then, we can get all the information for different $J_2/J_1$ 
from the result of a single simulation. Actually, for smaller 
system sizes ($L \le 16$), the two-parameter random walk method 
works well.  However, 
simulations with fixed $J_1$ and $J_2$ 
are more effective for larger sizes.  

In this paper, we have considered the staggered polarization field 
for the NNN interaction.  The effect of the F NNN interaction 
is also interesting.  In this case, the system has the same 
universality class as the six-state clock model, which yields 
two Berezinskii-Kosterlitz-Thouless transitions \cite{Bere71,Kost73}.  
The precise calculation of this problem has been quite recently 
performed by using the level-spectroscopy method \cite{Nomu95} 
based on the exact diagonalization of the transfer matrix \cite{Otsu05}.  
A complementary study of the AF Potts model with the F NNN 
interaction using the Monte Carlo 
method is highly needed, and this research is now in progress. 


\section*{Acknowledgments}

The authors wish to thank N. Kawashima, H. K. Lee, 
and C.-K. Hu for valuable discussions.  
This work was supported by a Grant-in-Aid for Scientific Research 
from the Japan Society for the Promotion of Science.
The computation of this work has been done using computer facilities 
of Tokyo Metropolitan University and those of 
the Supercomputer Center, Institute for Solid State Physics, 
University of Tokyo.


\newpage

\begin{table}
\caption{The list of $T_c$, $\nu$, $\beta/\nu$, and moment ratio 
at $T=T_c$ for two phase transitions. The temperature is 
represented in units of $T/J_1$.}
\label{table1}
\vspace{2mm}
\begin{ruledtabular}
\begin {tabular}  {lllll}
$J_2/J_1$ & $T_c$ & $\nu$ & $\beta/\nu$ & ratio \\
\hline
high temp. \\
~1/4 & 0.6720(2) & 0.83(1) & 0.105(4) & 1.120(6) \\
~1/2 & 0.9000(3) & 0.83(1) & 0.114(4) & 1.138(6) \\
~1   & 1.3150(5) & 0.83(1) & 0.122(4) & 1.148(6) \\
~2   & 2.1911(5) & 0.83(1) & 0.126(4) & 1.154(6) \\
~4   & 4.0840(5) & 0.82(1) & 0.127(4) & 1.154(6) \\
\hline
low temp. \\
~1/4 & 0.2830(2) & 1.00(1) & 0.120(4) & 1.162(4) \\
~1/2 & 0.5642(3) & 1.01(1) & 0.123(4) & 1.165(4) \\
~1   & 1.0657(5) & 1.01(1) & 0.123(4) & 1.164(4) \\
~2   & 1.8228(5) & 0.99(1) & 0.125(4) & 1.164(4) \\
~4   & 3.0240(5) & 0.98(1) & 0.123(4) & 1.155(4) \\
\end{tabular}
\end{ruledtabular}
\end{table}

\begin{table}
\caption{The estimate of the residual entropy per spin of 
the AF three-state Potts model on the square lattice.  
Two procedures are employed to calculate $S/N$. 
(1) The total number of states for the subspace of 
the ground states of the AF three-state Potts model is calculated. 
(2) The direct way to calculate the ground-state DOS 
of the three-state Potts model is used with the condition of 
$\sum_E g(E)=3^N$. 
}
\label{table2}
\vspace{2mm}
\begin{ruledtabular}
\begin {tabular}  {lll}
$L$ & $S(1)/N$ & $S(2)/N$ \\
\hline
16 & 0.43569(8) & 0.43575(11) \\
24 & 0.43345(6) & 0.43353(7)  \\
32 & 0.43259(4) & 0.43262(9)  \\
48 & 0.43198(3) & 0.43202(15) \\
64 & 0.43176(5) & 0.43185(10) 
\end{tabular}
\end{ruledtabular}
\end{table}


\begin{thebibliography}{99}

\bibitem{Pott52}
 Potts R B 1952 {\it Proc. Cambridge. Philos. Soc.} {\bf 48} 106 \\
 Wu F Y 1982 {\it Rev. Mod. Phys.} {\bf 54} 235

 \bibitem{Lena67}
 Lieb E H 1967 {\it Phys. Rev.} {\bf 162} 162 \\
 Lieb E H 1967 {\it Phys. Rev. Lett.} {\bf 18} 692

 \bibitem{Baxt70}
 Baxter R J 1970 {\it J. Math. Phys. (N.Y.)} {\bf 11} 3116 \\
 Baxter R J 1982 {\it Proc. Roy. Soc. London} {\bf A383} 43

 \bibitem{Nijs82}
 den Nijs M P M, Nightingale M P and Schick M 1982
 {\it Phys. Rev.} B {\bf 26} 2490

 \bibitem{Burt97}
 Burton Jr. J K and Henley C L 1997 
 {\it J. Phys. A: Math. Gen.} {\bf 30} 8385

 \bibitem{Wang89}
 Wang J S, Swendsen R H and Koteck\'y R 1989 
 {\it Phys. Rev. Lett.} {\bf 63} 109

 \bibitem{Sala98}
 Salas J and Sokal A D 1998 {\it J. Stat. Phys.} 
 {\bf 92} 729

 \bibitem{Ferr99}
 Ferreira S J and Sokal A D 1999 {\it J. Stat. Phys.} 
 {\bf 96} 461

 \bibitem{Card01}
 Cardy J, Jacobsen J L and Sokal A D 2001 {\it J. Stat. Phys.}
 {\bf 105} 25

 \bibitem{Otsu04}
 Otsuka H and Okabe Y 2004 {\it Phys. Rev. Lett.}
 {\bf 93} 120601

 \bibitem{DelfNATO}
 Delfino G, in 
 {\em Proceedings of the NATO advanced research workshop on Statistical
 field theories},
 edited by Cappelli A {\it et al.} (Kluwer A.P., 2002); eprint
 hep-th/0110181

\bibitem{berg91} 
 Berg B A and Neuhaus T 1991 {\it Phys. Lett.} B {\bf 267} 249 \\
 Berg B A and Neuhaus T 1992 {\it Phys. Rev. Lett.} {\bf 68} 9
\bibitem{Lee93}
 Lee J 1993 {\it Phys. Rev. Lett.} {\bf 71} 211

\bibitem{wl}
 Wang F and Landau D P 2001 {\it Phys. Rev. Lett.} {\bf 86} 2050 \\
 Wang F and Landau D P 2001 {\it Phys. Rev.} E {\bf 64} 056101

\bibitem{Binder} 
  Binder K 1981 {\it Z. Phys.} B {\bf 43} 119

\bibitem{Tome}
  Tom\'e T and Petri A 2002 {\it J. Phys. A: Math. Gen.} {\bf 35} 5379

\bibitem{Francesco}
 Di Francesco P, Saleur H and Suber J B 1987 
 {\it Nucl. Phys.} B {\bf 290} [FS20] 527 \\
 Di Francesco P, Saleur H and Suber J B 1988 
 {\it Europhys. Lett.} {\bf 5} 95

\bibitem{Zamo86} 
 Zamolodchikov A B 1986 
 {\it Zh. Eksp. Teor. Fiz.} {\bf 43} 565
 [1986 {\it JETP Lett.} {\bf 43} 730]

\bibitem{Banavar}
 Banavar J B, Grest G S and Jasnow D 
 1980 {\it Phys. Rev. Lett.} {\bf 45} 1424 \\
 Banavar J B, Grest G S and Jasnow D 
 1982 {\it Phys. Rev.} B {\bf 52} 4639

\bibitem{Yama01}
 Yamaguchi C and Okabe Y 2001 {\it J. Phys. A: Math. Gen.} {\bf 34} 8781

\bibitem{ent}
 Park H and Widom M 1989 {\it Phys. Rev. Lett.} {\bf 63} 1193

\bibitem{Shrock}
 Shrock R and Tsai S H 1997 {\it J. Phys. A: Math. Gen.} {\bf 30} 495 \\
 Shrock R and Tsai S H 1998 {\it Phys. Rev.} E {\bf 58} 4332

\bibitem{Cardy}
 Cardy J L 1996 {\it Scaling and Renormalization in Statistical Physics} 
 (Cambridge University Press, Cambridge)

\bibitem{Bere71}
 Berezinskii V L 1971 
 {\it Zh. Eksp. Teor. Fiz.} {\bf 61} 1144 
 [1972 {\it Sov. Phys. JETP} {\bf 34} 610]

\bibitem{Kost73}
 Kosterlitz J M and Thouless D J 1973
 {\it J. Phys. C: Solid State Phys.} {\bf 6} 1181 \\ 
 Kosterlitz J M 1974 {\it J. Phys. A: Math. Gen.} {\bf 7} 1046

\bibitem{Nomu95}
 Nomura K 1995 {\it J. Phys. C: Solid State Phys.} {\bf 28} 5451

\bibitem{Otsu05}
 Otsuka H, Mori K, Okabe Y and Nomura K 2005 {\it Phys. Rev.} E 
 {\bf 72} 046103

\end{thebibliography}
\end{document}